\documentclass[acmsmall]{acmart}
\usepackage{latexsym}
\usepackage{url}
\usepackage{CJKutf8}
\usepackage{multirow}
\usepackage{comment}
\usepackage{tabularx}
\usepackage{subfig}
\AtBeginDocument{%
  \providecommand\BibTeX{{%
    \normalfont B\kern-0.5em{\scshape i\kern-0.25em b}\kern-0.8em\TeX}}}

\begin{document}
\title{Studying Politeness across Cultures using English Twitter and Mandarin Weibo}
\author{Mingyang Li}
\email{myli@alumni.upenn.edu}
\orcid{0000-0002-8344-8462}
\affiliation{%
  \institution{University of Pennsylvania}
  \city{Philadelphia}
  \state{Pennsylvania}
  \postcode{19104}
}

\author{Louis Hickman}
\email{lchickma@purdue.edu}
\affiliation{%
  \institution{Purdue University}
  \streetaddress{610 Purdue Mall}
  \city{West Lafayette}
  \state{Indiana}
  \postcode{47907}
}

\author{Louis Tay}
\email{stay@purdue.edu}
\affiliation{%
  \institution{Purdue University}
  \streetaddress{610 Purdue Mall}
  \city{West Lafayette}
  \state{Indiana}
  \postcode{47907}
}

\author{Lyle Ungar}
\email{ungar@cis.upenn.edu}
\affiliation{%
  \institution{University of Pennsylvania}
  \city{Philadelphia}
  \state{Pennsylvania}
  \postcode{19104}
}

\author{Sharath Chandra Guntuku}
\email{sharathg@seas.upenn.edu}
\affiliation{%
  \institution{University of Pennsylvania}
  \city{Philadelphia}
  \state{Pennsylvania}
  \postcode{19104}
}
\renewcommand{\shortauthors}{Li, et al.}

\begin{abstract}
  Modeling politeness across cultures helps to improve intercultural communication by uncovering what is considered appropriate and polite. We study the linguistic features associated with politeness across American English and Mandarin Chinese. First, we annotate 5,300 Twitter posts from the United States (US) and 5,300 Sina Weibo posts from China for politeness scores.
  Next, we develop an English and Chinese politeness feature set, `PoliteLex'. Combining it with validated psycholinguistic dictionaries, we study the correlations between linguistic features and perceived politeness across cultures. We find that on Mandarin Weibo, future-focusing conversations, identifying with a group affiliation, and gratitude are considered more polite compared to English Twitter. Death-related taboo topics, use of pronouns (with the exception of honorifics), and informal language are associated with higher impoliteness on Mandarin Weibo than on English Twitter. Finally, we build language-based machine learning models to predict politeness with an F1 score of $0.886$ on Mandarin Weibo and $0.774$ on English Twitter.
\end{abstract}



\maketitle
\begin{CJK*}{UTF8}{gbsn}
\section{Introduction}
    Politeness is suggested to be a universal characteristic in that it is both a deeply ingrained value and also used to protect the desired public image, of the people involved in interactions~\cite{levinson_politeness:_1987}. Meanwhile, there are linguistic differences in the way politeness is expressed across cultures~\cite{janney1993universality,holtgraves1990politeness}. Past work on cross-cultural differences in politeness has primarily analyzed choreographed dialogs (such as scripts from Japanese TV dramas~\cite{matsumoto1989politeness}) or responses to fictional situations (\textit{e.g.} via a Discourse Completion Test~\cite{tawalbeh2012directness}). However, to better understand both the similarities and differences in politeness across cultures, it is important to study natural communication within each culture. This can serve to help in developing robust understanding of linguistic markers associated with politeness specific to each culture and across cultures. With people increasingly communicating on social media platforms, online posts can be used as a source of natural communication. While there is a large literature study trolling~\cite{viviani2017credibility} and hate-speech~\cite{schmidt2017survey} on social media, politeness is comparatively less explored.
    
    \paragraph{Practical importance of this study.}
    Practically, this work can also inform social media users on how to adopt more polite expressions--this is particularly important for people engaging in cross-cultural communication and writing in unfamiliar languages. There are cultural nuances to how politeness in language is expressed and perceived~\cite{chen1993responding}. For example, a generous invitation in Mandarin (\textit{e.g.} `来跟我们吃饭，
   不然我可生气了!' which literally translates to `Come and eat with us, or I will get mad at you!') may sound utterly rude to a typical American~\cite{gu1990politeness}. 
    
    \paragraph{Politeness theory.}
    To study (im-)politeness, several frameworks have been proposed from a pragmatic viewpoint~\cite{kadar2017politeness}. In 1987, Brown and Levinson identified several types of politeness strategies based on face theory~\cite{levinson_politeness:_1987}. Later, Leech developed a maxim-based theory that accommodates more cultural variations~\cite{leech2007politeness}, which is arguably the most suitable for East-West comparison~\cite{al-duleimi2016acritical}. Culture can play a significant role in shaping emotional life. Specifically, different cultures may value different types of emotions (e.g., Americans value excitement while Asians value calm) \cite{tsai2006cultural}, and there are different emotional display rules across cultures \cite{matsumoto1990cultural}. In general, psychological research reveals both cultural similarities and differences in emotions \cite{elfenbein2003universals}. We take inspiration from these theories for developing computational tools to study cross-cultural (im-)politeness. 
    
    \paragraph{Prior work.}
    A seminal computational linguistics study on politeness~\cite{danescu-niculescu-mizil2013acomputational} has developed framework for predicting politeness in English requests, sourced from Wikipedia and Stack Exchange, using bi-grams and semantic features derived from parse trees. The Stanford Politeness corpus introduced in the same study has the following characteristics:
    i) posts are taken from forums where power hierarchy between users and moderators is explicit; 
    ii) each post in the corpus contains exactly two sentences and always ends in a question mark; and, most importantly for cross-lingual research,
    iii) some features do not apply to other languages.
    
    \paragraph{Research questions and contributions.}
	We introduce a microblog corpora consisting of $5,300$ posts from US English Twitter and $5,300$ posts from Chinese Sina Weibo, each annotated with politeness score, in order to compare how (im-)politeness is expressed by speakers and understood by readers in each language. We design a politeness feature set (`PoliteLex') for English and Mandarin.  By building a politeness-specific lexicon as our feature set, we are able to ensure a degree of linguistic equivalence across cultures for comparison~\cite{bargiela-chiappini_politeness_2010,chen1993responding}. Modeling politeness across cultures can also inform organizations, expatriates, and machine translation tools by elucidating how to make text-based speech acts appropriate and polite for the listener's culture, rather than the speaker's culture. Adjusting communication to be culturally appropriate is necessary to avoid misunderstandings and improve cross-cultural relations~\cite{thomas1984cross}. To these ends, we aim to answer the following questions:
    \begin{enumerate}
    	\item How do psycho-linguistic lexica (including PoliteLex) compare to the Stanford API~\cite{danescu-niculescu-mizil2013acomputational} at predicting (im-)politeness on the Stanford Politeness Corpus built on Wikipedia and StackExchange posts?
    	\item What are the differences (and similarities) in politeness expressions between US English and Mandarin Chinese (represented by Twitter and Sina Weibo, respectively)?
    	\item How accurately can machine learning algorithms predict politeness in microblog corpus in both US English and Mandarin Chinese?
    \end{enumerate}

    \paragraph{Advantages of using lexica.}
    In addition to PoliteLex, we also employ Linguistic Inquiry Word Count (LIWC)~\cite{pennebaker2001linguistic} and NRC Word-Emotion Association Lexicon (EmoLex)~\cite{Mohammad13} in our study. This lexica-based approach i) allows us to compare politeness expressions across languages and ii) generates insights that map onto known psycho-social aspects across the languages. Then, we examine similarities and differences in politeness expressions across languages (and potentially cultures) using PoliteLex, LIWC, and EmoLex with microblog corpora. 
    
    \paragraph{Advantages of using microblog corpora.}
    Microblogs have the advantage of capturing communication in a natural online setting and also provide better cultural representation compared to request forums. Microblog texts are expected to be less task-oriented and less pragmatically constrained compared to collaborative platforms such as Wikipedia and Stack Exchange. This suggests that a prediction model built upon microblog corpora may generalize better than the Stanford API.
    Second, Twitter and Weibo have relatively isolated user bases. We retain English Twitter posts within the United States (US) and Mandarin Weibo posts within China to further avoid bilingual confounds. Moreover, microblogs could be an excellent avenue for studying politeness as a majority of the people get their news and daily updates on microblogs~\cite{kothari2013detecting}, and resources to aid polite conversations could be beneficial in several downstream analysis (eg: reducing polarization~\cite{tucker2018social}). 
    
    \paragraph{Findings summary.}
    Our findings can be summarized as follows: (1) features derived from psycho-linguistic lexica outperform bi-grams and semantic features~\cite{danescu-niculescu-mizil2013acomputational} in predicting (im-)politeness in the Stanford Politeness Corpus ($3\%$ increase in F-1 score). (2) On Mandarin Weibo, future-focusing conversations, identifying with a group affiliation, and gratitude are considered to be more polite compared to English Twitter. And death-related taboo topics, lack or poor choice of pronouns, and informal language are associated with higher impoliteness on Mandarin Weibo compared to English Twitter. (3) Machine learning models trained on language features can predict (im-)politeness on English Twitter with an F1-score of $0.774$ and on Mandarin Weibo with an F1-score of $0.886$. \footnote{PoliteLex, code, and pre-trained politeness models are available at \url{https://github.com/tslmy/politeness-estimator}.}.
    
    \paragraph{Paper organization.}
    In the next section, we introduce PoliteLex, describe some of the categories in this lexicon before using it to predict politeness on an existing politeness corpus from~\cite{danescu-niculescu-mizil2013acomputational}. We then describe the Weibo Twitter Politeness Corpus, consisting of annotated English Twitter posts from the US and Mandarin Sina Weibo posts from China. Next, we study linguistic differences in politeness expressions across the two both languages. Finally, we build predictive models of politeness in both languages using several language features.

\section{PoliteLex: Politeness Features for English and Mandarin}

    
    Curated lexica have long been employed in psycholinguistic research~\cite{john_lexical_1988}. A lexicon is often a many-to-many mapping of tokens (including words and word stems) to categories. For example, LIWC~\cite{pennebaker2001linguistic} maps tokens into psychological categories, and EmoLex~\cite{Mohammad13} consists of $14,182$ English unigrams mapped to $7$ emotion categories -- joy, surprise, anticipation, sadness, fear, anger, and disgust.

    Analogous to LIWC and EmoLex, we develop PoliteLex, a lexicon that maps tokens (or regular expressions when it is required to anchor to the beginning of text) to (im-)politeness markers.
	In PoliteLex, tokens are curated from various sources to capture semantic (such as \texttt{hedges}, \texttt{ingroup\_ident}, and \texttt{taboo}) as well as syntactical features (such as \texttt{could\_you}, \texttt{can\_you}, \texttt{start\_i}, and \texttt{first\_person\_plural}). While several PoliteLex features (listed in Table \ref{tab:feats}) are inspired from the semantic features used by prior work~\cite{danescu-niculescu-mizil2013acomputational} in English, appropriate modifications (including addition of new features) were made to make the lexicon compatible with Mandarin. We identify 18 features from prior work~\cite{danescu-niculescu-mizil2013acomputational} with strategies introduced in existing (im-)politeness theories and add 8 new features for strategies that are not covered by prior work~\cite{danescu-niculescu-mizil2013acomputational}. For example, the \texttt{praise} and \texttt{gratitude} features align with Leech's approbation maxim, which involves expressing approval of the hearer~\cite{leech2007politeness}. Conversely, some of the features, like \texttt{please} and \texttt{start\_with\_please}, represent impoliteness strategies when withheld since the hearer may be expecting these expressions. Importantly, understanding politeness expressions requires semantic understanding that goes beyond isolated words and phrases, which is why a top-down, theoretically-driven approach to developing PoliteLex needs to be paired with the bottom-up, data-driven approach of uncovering the context-specific relationships between the features and (im-)politeness.

    As shown in Table \ref{tab:feats}, the Chinese version and the English version of PoliteLex are designed with identical feature names. However, this does not imply either version of PoliteLex to be a literal translation from the other. Instead, tokens are chosen based on whether they appropriately represent the linguistic trait in the respective language according to prior studies, e.g., \cite{leech2014thepragmatics,owen1983apologies, GUAN200932,song-cen1991socialdistribution,hong1985politeness,chinasafari,erbaugh2008chinaexpands,zhou2012astudy}. In other words, PoliteLex uses politeness theory and cultural knowledge to derive culture-specific markers of (im-)politeness. Some of the features include:

    \begin{table*}[t!]
            \footnotesize
            \begin{centering}
                \begin{tabular}{|l|l|l|l|} 
                \hline
                \textbf{Description} & \begin{tabular}{l}\textbf{Corresponding} \\\textbf{Feature in~\cite{danescu-niculescu-mizil2013acomputational}}\end{tabular} & \textbf{Examples}                         & \textbf{Name}             \\ 
                \hline\hline
                apologetic                                    & apologizing                                                                                                  & my bad, sorry, 对不起               & \texttt{apologetic}               \\ 
                \hline
                honorific ``you''                               & /                                                                                                            & your honor, ur majesty, 您        & \texttt{you\_honorific}           \\ 
                \hline
                direct ``you''                                  & 2nd person                                                                                                   & you, u, 你                        & \texttt{you\_direct}              \\ 
                \hline
                hedges                                        & hedges                                                                                                       & doubtful, imo, 也许, 没准儿 & \texttt{hedge}                    \\ 
                \hline
                gratitude                                     & gratitude                                                                                                    & thanks, thx, 谢谢, 鸣谢, 感谢, 重谢      & \texttt{gratitude}                \\ 
                \hline
                taboo                                         & /                                                                                                            & dammit, fuck, 他妈的                & \texttt{taboo}                    \\ 
                \hline
                best wishes                                   & /                                                                                                            & have a great day, 您好             & \texttt{best\_wishes}             \\ 
                \hline
                praise                                        & deference                                                                                                    & awesome, bravo, 真棒               & \texttt{praise}                   \\ 
                \hline
                by-the-way                                    & indirectbtw                                                                                                  & by the way, btw, 对了, 说起来, 话说     & \texttt{indirect\_btw}            \\ 
                \hline
                please                                        & please                                                                                                       & please, pls, plz, 请              & \texttt{please}                   \\ 
                \hline
                start with please                             & pleasestart                                                                                                  & please, 请                        & \texttt{start\_please}            \\ 
                \hline
                emergency                                     & /                                                                                                            & asap, right now, 立刻, 马上          & \texttt{emergency}                \\ 
                \hline
                honorifics                                    & /                                                                                                            & Mr., Prof.,~令尊, 令堂, 令兄           & \texttt{honorifics}               \\ 
                \hline
                greeting                                      & greeting                                                                                                     & Hey, Hi, Hello, 嗨, 哈喽, 哈罗, 嘿     & \texttt{greeting}                 \\ 
                \hline
                promise                                       & /                                                                                                            & i promise, must, surely, 肯定, 绝对  & \texttt{promise}                  \\ 
                \hline
                start with so                                 & directstart                                                                                                  & So,~那                            & \texttt{start\_so}                \\ 
                \hline
                factuality                                    & factuality                                                                                                   & in fact, actually,~其实, 说实话, 讲真   & \texttt{factuality}               \\ 
                \hline
                counterfactual modal                          & could                                                                                                        & could you, would u, 你想不想   & \texttt{could\_you}               \\ 
                \hline
                indicative modal                              & can                                                                                                          & can you, will u, 你可…吗？   & \texttt{can\_you}                 \\ 
                \hline
                start with question                           & direct                                                                                                       & what, why, 为什么, 怎, 咋             & \texttt{start\_question}          \\ 
                \hline
                in-group identity                             & /                                                                                                            & mate, bro, homie,~咱,~咱们          & \texttt{ingroup\_ident}           \\ 
                \hline
                first person plural                           & 1pl                                                                                                          & we, our, us, ours, 我们            & \texttt{first\_person\_plural}    \\ 
                \hline
                first person singular                         & 1                                                                                                            & i, my, mine, me, 我, 俺            & \texttt{first\_person\_singular}  \\ 
                \hline
                together                                      & /                                                                                                            & together, 一起, 一同                 & \texttt{together}                 \\ 
                \hline
                start with i                                  & 1start                                                                                                       & i, 我                             & \texttt{start\_i}                 \\ 
                \hline
                start with you                                & 1start                                                                                                       & you, u, 你                        & \texttt{start\_you}               \\
                \hline
                \end{tabular}
            \par\end{centering}
            \caption{\label{tab:feats} List of lexical features considered in PoliteLex. The features that anchor to the beginning of sentences employ regular expressions. } 
        \end{table*}

    \paragraph{The magic word `please' and its position in sentence.}
    Prior work~\cite{danescu-niculescu-mizil2013acomputational} indicates that, while the word `please' is generally regarded as a definitive politeness marker, a please-starting sentence may sound imperative and thus less polite compared to a mid-sentence `please'. To validate this hypothesis, we design a feature \texttt{please} that captures the word regardless of position, and \texttt{start\_please} capturing the sentence-leading case. Variants such as `plz' and `pls', possibly with `s' or `z' repeated, are also considered.
    \paragraph{Apologies.}
    Apology is the speech employed in an attempt to redress a previous transgression~\cite{leech2014thepragmatics,owen1983apologies}. In feature \texttt{apologetic}, we capture expressions such as `sorry', `sry', `my bad', `my fault', etc., in English and `对不起', `抱歉', `不好意思', `很遗憾', etc., in Chinese based on prior work~\cite{GUAN200932}.
    \paragraph{Greetings and wishes.}
    Besides apologies, greetings are also very culture-specific pragmatic particles that convey politeness. For example, `你好' (nĭ hăo), the everyday greeting phrase in Mandarin Chinese, is often translated to `how are you'~\cite{song-cen1991socialdistribution,hong1985politeness} or simply `hello'~\cite{chinasafari,erbaugh2008chinaexpands}, while a more literal translation should be `you (doing) well'~\cite{zhou2012astudy}, conveying a wish of good life, good health, or good day for the hearer. To capture this non-phatic implication, we designed two separate features, \texttt{best\_wishes} and \texttt{greetings}. To ensure comparability across languages, we only capture the phrases `wish/hope you/everyone/y'all' in \texttt{best\_wishes}, and in \texttt{greetings} we capture `hi', `hey', and `hello'.
    \paragraph{Maxim-based features.}
    Several features are designed to capture markers that correspond to politeness maxims in Leech's theory~\cite{leech2014thepragmatics}. Feature \texttt{gratitude} attempts to capture speakers' recognition, approval, and appreciation towards hearers' work, corresponding to the approbation maxim. Feature \texttt{honorifics} targets the same maxim, searching for titles such as `Mr.' and `Prof.' in English and `your honorable father', etc., in Chinese. For the tact maxim, the feature \texttt{hedge} is designed to seek for tokens that mitigate intensities~\cite{kaltenbock2010newapproaches}.
    Since aspects such as topic and emotion also affects expression of politeness besides mere markers described above, PoliteLex is designed as a complementary tool to other lexica. In the follow sections, we employ a combination of LIWC, EmoLex, and PoliteLex for studying (im-)politeness.

    \section{RQ1: Predicting Politeness using Lexica on Stanford Politeness Corpus}
    
    \paragraph{The Stanford Politeness Corpus.} Posts are collected from StackExchange and Wikipedia Talk pages~\cite{danescu-niculescu-mizil2013acomputational}. Using Amazon Mechanical Turk, each post is rated for politeness by $5$ independent annotators, who are requested to annotate $13$ posts per batch. A total of $10,956$ posts are obtained. Next, only the top-rated $25\%$ posts (labeled as `polite') and the lowest $25\%$ (labeled as `rude') are kept for training and testing. There are $2,739$ posts in each class. $500$ posts are sampled from the two classes for testing.

    \paragraph{Predictive modeling.}
    We train prediction models on the Stanford Politeness Corpus with features used in their work \cite{chang2020convokit}, which consist of $1,426$ bigrams and $21$ parse-tree-based rules, and various combinations of the lexical features defined by PoliteLex, LIWC, and EmoLex.
    Support Vector Machine Classifiers (SVC) are trained on the $4,978$ posts with linear kernel, $L2$ regularization penalty parameter of $C=0.02$, and stopping tolerance of $\epsilon=10^{-3}$. These choices are made to enable apples-to-apples comparison with prior work~\cite{danescu-niculescu-mizil2013acomputational}. 
    
    Prediction performances are shown in Table \ref{tab:polipredperf}. The model trained solely with the LIWC features achieves similar performance to the one trained on the API features. PoliteLex performs only slightly worse than LIWC and Stanford API, although it was developed using communication from a different context. Combination of LIWC, EmoLex, and PoliteLex achieves the best performance on the Stanford Politeness Corpus.

    \begin{table*}[t!]
        \small
        \begin{centering}
            \begin{tabular}{|l|r|r|r|r|r|}
            \hline
            \textbf{Feature Set}                         &          \textbf{F1} &    \textbf{Precision} &    \textbf{Recall} &    \textbf{ROCAUC} &    \textbf{Accuracy}  \tabularnewline\hline\hline
            Stanford Politeness API             &     $0.663$ &      $0.686$ &   $0.672$ &   $0.668$ &     $0.672$  \tabularnewline\hline
            LIWC Only                           &     $0.670$ &      $0.673$ &   $0.672$ &   $0.670$ &     $0.672$  \tabularnewline\hline
            EmoLex Only                         &     $0.488$ &      $0.568$ &   $0.534$ &   $0.542$ &     $0.534$  \tabularnewline\hline
            PoliteLex Only            &     $0.603$ &      $0.626$ &   $0.616$ &   $0.611$ &     $0.616$  \tabularnewline\hline
            LIWC + EmoLex                       &     $0.659$ &      $0.660$ &   $0.660$ &   $0.659$ &     $0.660$  \tabularnewline\hline
            LIWC + PoliteLex          &     $0.676$ &      $0.679$ &   $0.678$ &   $0.676$ &     $0.678$  \tabularnewline\hline
            LIWC + EmoLex + PoliteLex &     $\textbf{0.685}$ &      $\textbf{0.687}$ &   $\textbf{0.686}$ &   $\textbf{0.684}$ &     $\textbf{
0.686}$  \tabularnewline\hline
            
            \end{tabular}
        \par\end{centering}
        \caption{\label{tab:polipredperf}Politeness Prediction Performance of lexica and the Stanford API evaluated on the Stanford Politeness Corpus. Highest value in each column is marked in bold.}
    \end{table*}

\section{RQ2: Similarities and Differences in Politeness Expressions in US and China}

    
    \subsection{Weibo Twitter Politeness Corpus}
        \paragraph{Collecting microblog posts.}
        
            Twitter posts are obtained from a 10\% archive released by the \textit{TrendMiner} project~\cite{preotiuc2012trendminer}, and Weibo posts are obtained using a breadth-first search strategy on Weibo users~\cite{guntuku2019studying}. On Twitter, the coordinates or tweet country locations (whichever was available) are used to geo-locate posts~\cite{schwartz2013characterizing}. On Weibo, user’s self-identified profile locations are used to identify the geo-location of messages. We use messages posted in the year 2014 in both corpora. To remove the confounds of bilingualism~\cite{fishman1980bilingualism}, we filter posts by the languages they are composed in and from US Twitter and Sina Weibo in China. Language used in each post is detected via \textit{langid}~\cite{lui2012langid}. After discarding direct re-tweets, $5,300$ posts are sampled from each platform. This study received approval from authors' Institutional Review Board (IRB).

        \paragraph{Annotating posts for politeness.}
        
            Separately on both Twitter and Weibo corpora of $5,300$ posts each, two native language annotators independently label the posts for perceived politeness on a integer scale of $-3\sim+3$, from most impolite to most polite. Prior to annotation, one of the paper's authors met with the annotators to review a list of politeness strategies~\cite{danescu-niculescu-mizil2013acomputational}, their definitions, and examples from social media to provide a frame of reference for the annotation process. On (Twitter, Weibo), politeness annotations have min=($-3$, $-3$), max=($3$, $3$), mean=($.137$, $-.041$), and median=($.333$, $0$), respectively.
            We standardize the scores within each annotator and average across them to obtain the final rating for each post.
            The inter-rater reliability (Krippendorff's alpha, interval) is $.528$ for Twitter, and $.661$ for Weibo.
                    \begin{figure}[h!]
            	\begin{centering}
            		\includegraphics[width=.5\columnwidth]{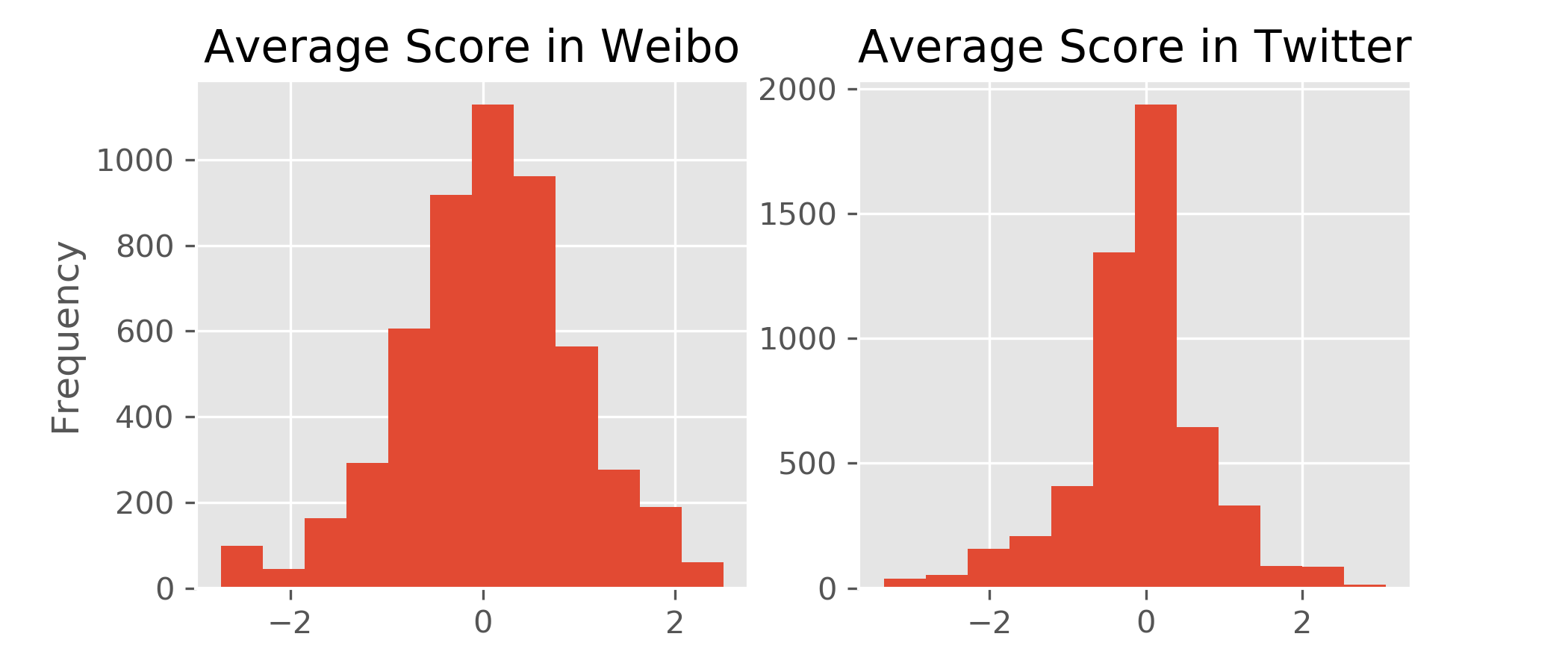}
            		\par\end{centering}
            	\caption{Annotated politeness scores in each microblog corpus, averaged across annotators.\label{fig:Histogram-of-politeness} } 
            \end{figure}
            
            The data collection effort took place in two batches. On the Twitter corpus, two annotators label $2,985$ posts, while another two label $2,940$ posts. $625$ posts in the $2,940$ batch are sampled from the previous $2,985$ batch considering the lower alpha value on Twitter compared to Weibo. The $625$ posts achieve a ICC(2,k) of $0.766$ across all $4$ annotators. In subsequent analysis, we use unique $5,300$ posts from each corpora.

    \subsection{Extracting Features}
        
        Twitter text is tokenized using \textit{Social Tokenizer} bundled
        with \textit{ekphrasis}~\cite{baziotis2017datastories}, a text processing
        pipeline designed for social networks. Using \textit{ekphrasis}, URLs,
        email addresses, percentages, currency amounts, phone numbers, user
        names, emoticons, and time-and-dates are normalized with meta-tokens
        such as `\texttt{<url>}', `\texttt{<email>}', `\texttt{<user>}' etc. Weibo posts are segmented using \textit{jieba}\footnote{\url{https://github.com/fxsjy/jieba}}
        considering its similar ability to handle highly colloquial corpus such as Sina Weibo.
        We vectorize each post by counting tokens in the post against the token list, representing percentage proportions of each feature --  Linguistic Inquiry Word Count (LIWC), NRC Word-Emotion Association Lexicon (EmoLex), and PoliteLex. 

    \subsection{Correlational Analysis}

\begin{figure}[t!]
            \begin{centering}
                \includegraphics[width=.75\columnwidth]{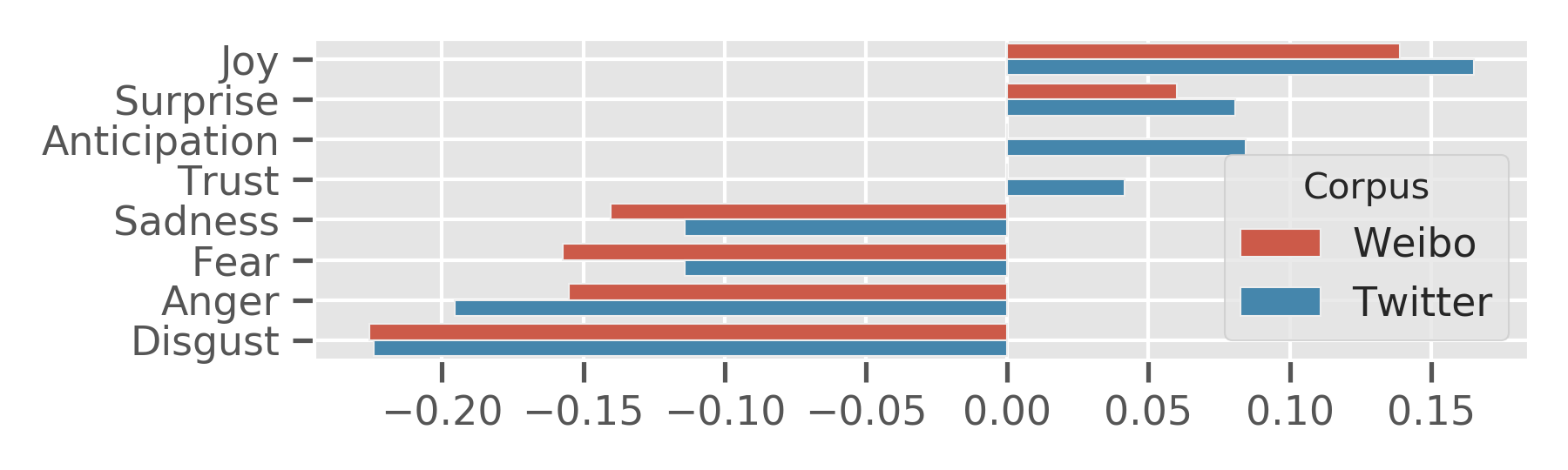}
            \par\end{centering}
            \caption{\label{fig:pcc-emolex} Pearson correlations between EmoLex
            and politeness scores on both platforms. Insignificant correlations ($p>.01$) are omitted.}
        \end{figure}

\begin{figure}[t!]
            \begin{centering}
                \includegraphics[width=.75\columnwidth]{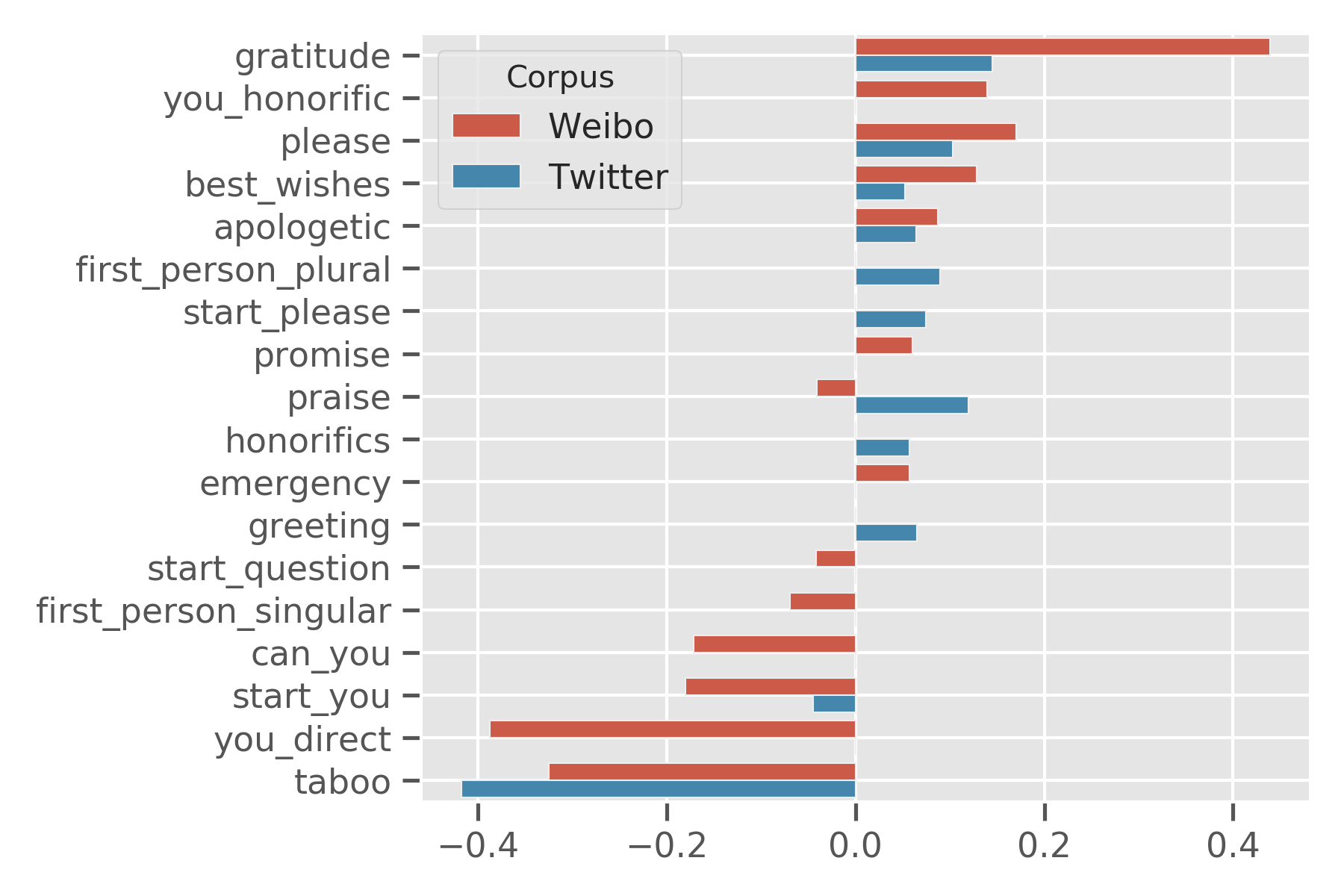}
            \par\end{centering}
            \caption{\label{fig:pcc-add} Pearson correlations between PoliteLex
            and politeness scores on both platforms. Insignificant  correlations ($p>.01$) are omitted.}
        \end{figure}

\begin{figure}[t!]
        	\begin{centering}
        		\subfloat{\includegraphics[width=.5\columnwidth]{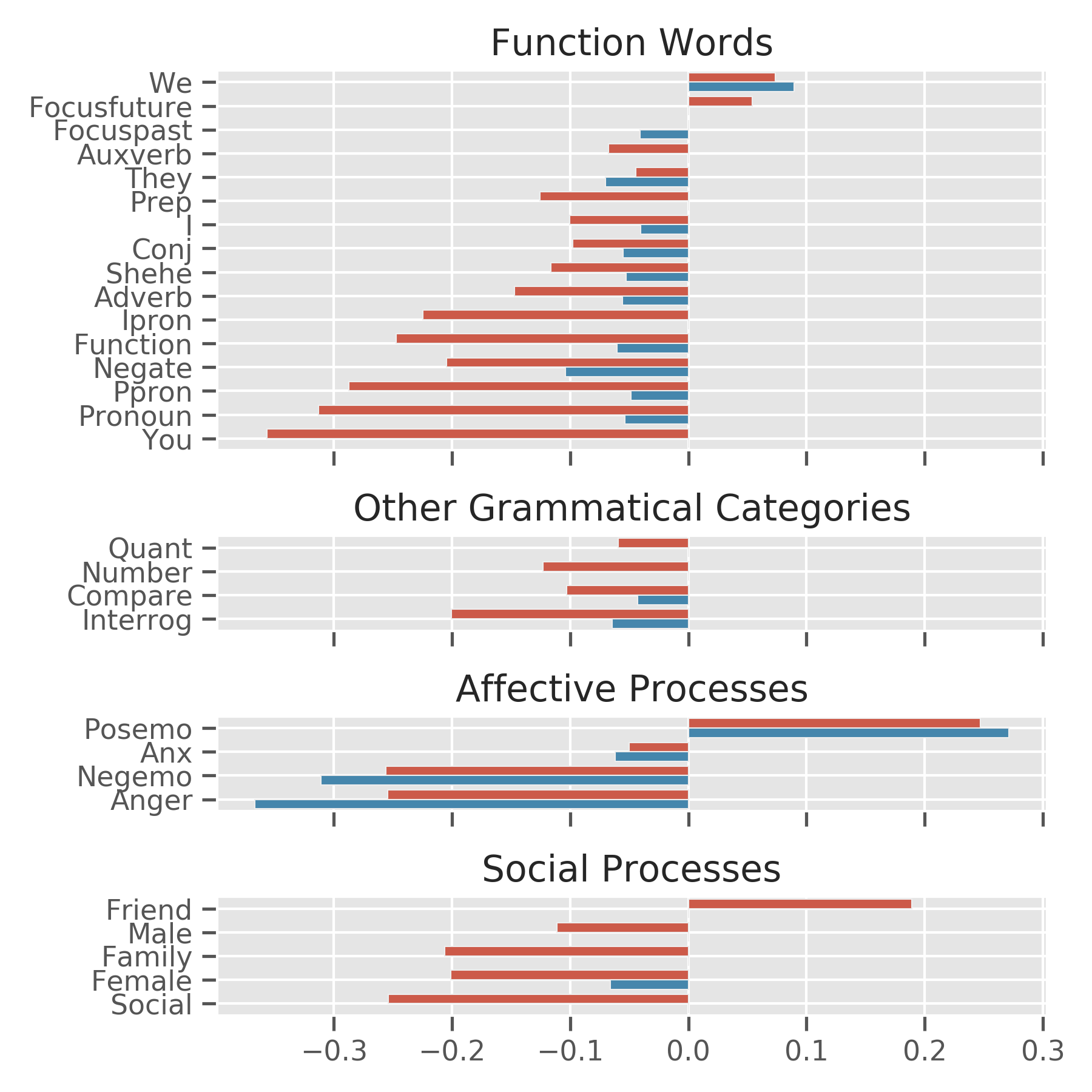}}%
        		\subfloat{\includegraphics[width=.5\columnwidth]{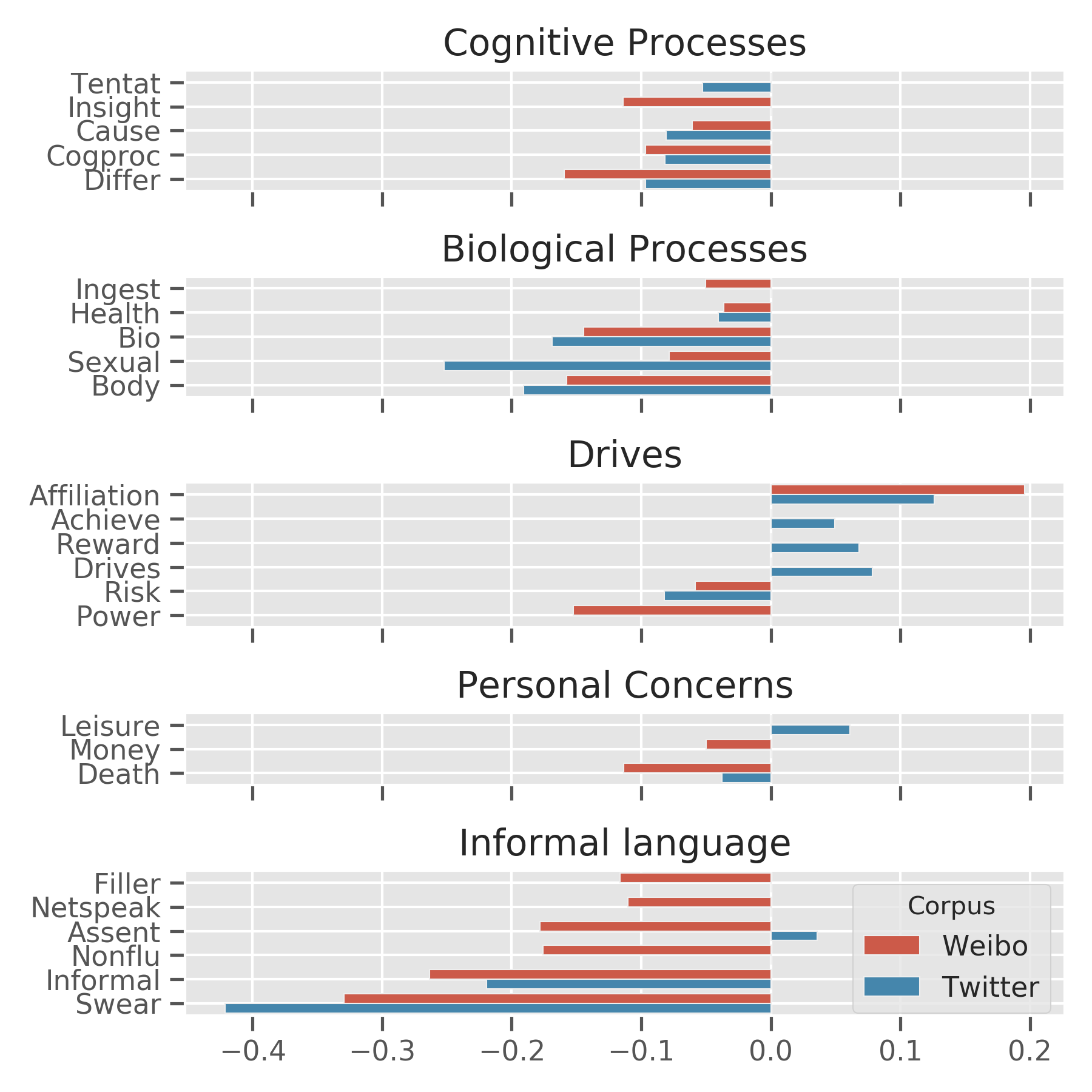}}
        		\par\end{centering}
        	\caption{\label{fig:pcc-liwc} Pearson correlations between LIWC categories
        		and average politeness scores on both platforms. Insignificant correlations ($p>0.01$) are omitted.
        	}
        \end{figure}

        In each corpus of Twitter and Weibo, for each feature, we compute its Pearson Correlation Coefficient (PCC) with the average scores across all $5,300$ posts. We consider correlations significant if they are less than a Bonferroni-corrected $p < 0.01$. Figure \ref{fig:pcc-emolex}, Figure \ref{fig:pcc-add}, and Figure \ref{fig:pcc-liwc} show the PCCs of Emolex, PoliteLex, and LIWC respectively. Features whose correlations are not significant ($p>0.01$) on both platforms are omitted. Of the 26 PoliteLex features, 18 are significantly correlated with (im-)politeness on either Weibo or Twitter.

        Studying the correlations between different lexical categories and politeness across Weibo and Twitter, we find that several map onto known psycho-social differences (and similarities) in US and Chinese cultures.
        
        \paragraph{Affect and emotions.} 
        Negative and positive emotions (\texttt{Negemo} and \texttt{Posemo}, in Figure \ref{fig:pcc-liwc}), correlate with impoliteness and politeness, respectively. General affective states tend to have similar relationships with (im-)politeness on Weibo and Twitter, suggesting the potential universality of (im-)politeness with respect to emotions~\cite{janney_universality_2009}. 
        When specific emotions are examined by LIWC and EmoLex (see \texttt{Affective Processes} in Figure \ref{fig:pcc-liwc} and Figure \ref{fig:pcc-emolex}), some cultural differences do emerge. 
        The differences share a pattern: most emotions correlate more strongly on English Twitter than on Mandarin Weibo. This may be explained by the tendency in the collectivist culture of China to conceal one's emotions, such as `anger'~\cite{liu2014chinese}, because it disrupts social harmony, or \textit{guanxi}. One exception to this pattern is that \texttt{Fear} correlated more strongly with impoliteness on Weibo than on Twitter. Some studies have found that Chinese youth self-report higher levels of fear than Americans ~\cite{ollendick1996fears}, and Weibo users who fail to conceal these emotions may be viewed as highly impolite.
        In addition, \texttt{Disgust} has the closest PCCs across the two cultures among all emotions in Figure \ref{fig:pcc-emolex}. Impolite behavior may not only trigger the emotion `disgust' from the reader, but also encourage similarly impolite response that reflects the emotion `disgust'~\cite{Vogel2014}, which may explain this similarity in correlations.
        
        \paragraph{Gratitude.}
        One way of restoring \textit{guanxi} is via expressing gratitude. \texttt{Gratitude} is much more predictive of politeness on Weibo than on Twitter. Gratitude motivates and maintains social reciprocity which is much more important among collectivist societies, like China, than individualistic societies, like the US~\cite{floyd2018universals,wang_expressions_2015}. This is reflected in Figure \ref{fig:pcc-add}, where the PCC for \texttt{gratitude} in Chinese is almost triple of that in English. Gratitude fosters reciprocity \cite{algoe2008beyond}, which is even more necessary for social functioning in collectivist societies where individual goals tend to closely align with group goals.

        \paragraph{Emergency.}
        Under emergency, speakers are generally expected to (and excused of) omitting phatic expressions such as politeness markers. Polite conversations under emergency can be imperatives that are beneficial to the addressee (`Watch out!') or requests where the addressee are expected to help with the issue (`Help!'). Both situations are defined as expressions of optimism in prior studies~\cite{zhan1992strategies}. Other study suggests that Chinese speakers tend to expect less politeness from optimistic expressions compared to English speakers~\cite{chen2014being}. This is confirmed with our findings: when \texttt{emergency} markers exist, the perceived politeness correlates higher among Chinese speakers.

        Additionally, some \texttt{emergency} tokens may express responsiveness. In Example 1 of Table \ref{tab:weibo-examples}, the speaker demonstrates their gratitude by indicating their eagerness to download the file that that was shared. In this particular sense, \texttt{emergency} is related to \texttt{Focusfuture}, which is discussed below.

        \paragraph{The word `please'.}
        As introduced earlier, prior research claims that a sentence-starting `please' does not sound as polite as a mid-sentence `please'~\cite{danescu-niculescu-mizil2013acomputational}. Indeed, we found that the PCC of \texttt{start\_please} is lower than that of \texttt{please} (which captures both leading and mid-sentence `please') in both languages. 

        \paragraph{Informal language.}
        Most linguistic features in informal language (Figure \ref{fig:pcc-liwc}) are associated with impoliteness on Weibo more than Twitter. One way cultures differ is the strength of societal norms and tolerance of deviant behavior--when norms are strong and little deviance is allowed, cultures are considered tight; when norms are relatively weak and deviant behavior is tolerated, cultures are considered loose. The tightness or looseness of a culture emerges in response to challenges faced by the particular culture, and China has a tighter culture than the United States~\cite{gelfand_differences_2011}. Similarly, China's culture is characterized by relatively higher power distance ~\cite{hofstede2009geert}, necessitating formal language both up and down social hierarchy to maintain the hierarchical structure. Deviation from these norms is considered impolite in such cultures, reflected on Weibo in the impoliteness of using filler words, netspeak, and nonfluencies in general. 
        
        \paragraph{Taboo and swearing.}
        Death is a major taboo topic in Chinese culture~\cite{hsu2009understandings}, to an extreme that Chinese people avoid gifting timepieces because `gifting a clock' (送钟) in Chinese sounds identical to `giving terminal care' (送终)~\cite{xu2007death}. This avoidance of death-related vocabulary is reflected in our analysis: The feature \texttt{Death} correlates more with impoliteness on Weibo compared to Twitter. 
       \texttt{Swear} words are associated with impoliteness in both English Twitter and Mandarin Weibo. 
        In Brown and Levinson's terms, swearing can damage hearer's want to be treated with dignity and grace, constituting impoliteness~\cite{CULPEPER1996349}. 
        We also include a crowd-sourced list of taboo words in each language as a catch-all \texttt{taboo} feature (Figure \ref{fig:pcc-add}) that captures swearwords, death affairs, profanities, sexual language, etc. This feature performs better at capturing impoliteness on the Twitter side. 

        \paragraph{Pronouns.}
        The pronoun `you' can be translated to two forms in Chinese: a honorific form, `nín' (您), and a direct form `nǐ' (你). With this subtle difference in mind, we capture the honorific form and the direct form with separate features, \textit{i.e.} \texttt{you\_honorific} and \texttt{you\_direct}. Indeed, it is observed that \texttt{you\_honorific} correlates with politeness while \texttt{you\_direct} correlates with impoliteness (see Example 2 in Table \ref{tab:weibo-examples}).
        However, the mere mention of `you', regardless of form, seems to imply impoliteness on Weibo but not on Twitter. This observation can be explained by prior findings that, in Chinese culture, explicit usage of the pronoun `you' may sound interrogative by imposing responsibility on the hearer's opinion~\cite{kadar2011politeness}. Supporting this rationale, direct instances of \texttt{Interrog} are associated with impoliteness. 
        In fact, since Mandarin speakers often omit pronouns in everyday conversation~\cite{hsiao2014agent}, all types of \texttt{Pronoun}, either personal (\texttt{Ppron}) or impersonal (\texttt{Ipron}), correlate with impoliteness in Weibo. This is also reflected in several of the Affective Processes categories as they also contain pronouns. Additionally, other Function Words may similarly be omitted in everyday Mandarin, leading them to be associated with impoliteness.
        
        \paragraph{In-group markers.}
        Claiming solidarity between the hearer and the speaker helps to find a common ground on which cooperation can happen, constituting positive politeness~\cite{levinson_politeness:_1987}.  
        On the other hand, suggesting difference between the hearer and the speaker may disrupt the common ground, making the speech sound impolite.
        Indeed, the feature \texttt{Affiliation} correlates with politeness, while the feature \texttt{Differ} correlates with impoliteness.
        Interestingly, both features correlate more significantly on Weibo than they do on Twitter. This agrees with previous comparative studies on collectivism and individualism between China and the US, which suggest that Chinese people are more sensitive about social harmony, the aforementioned \textit{guanxi}~\cite{FLAIRS124402}.
        
        Another related LIWC feature is \texttt{Friend}. In Example 3 of Table \ref{tab:weibo-examples}, the word `friends' can refer to `anyone', instead of a particular group of companions of the speaker. Nevertheless, the speaker chose to use `friends' instead of `anyone' to show politeness towards people who are in doubt by including them in the speaker's friend circle.
        
              \begin{table*}[h]
        	\small
        	\begin{centering}
        		\begin{tabularx}{\textwidth}{|l|X|p{12em}|}
        			\hline
        			\textbf{ID} & \textbf{Text} & \textbf{Features} \\ \hline\hline
        			1  & 谢谢分享！\textbf{马上}下来试试 / Thanks for sharing! (I'll) \textbf{immediately} download and try & \texttt{emergency}, \texttt{Gratitude}   \\ \hline
        			2 & \textbf{你}有孩子吗？！\textbf{你}知道孩子在有便意时无法自我控制吗？！ / (Do) \textbf{you} have kids?! (Do) \textbf{you} know that children cannot control themselves when they have a poop?! & \texttt{you\_direct}, \texttt{You}, \texttt{start\_you}            \\ \hline
        			3  & 我们敞开大门欢迎各位有质疑的\textbf{朋友}！      / We keep our doors open for \textbf{friends} in doubt!                                                   & \texttt{Friend}                                  \\ \hline
        			4  & @user thanks \textbf{dawgggg} & \texttt{Friend}, \texttt{Gratitude}                                 \\ \hline
        			5  & 实在很抱歉。\textbf{以后}注意。    / (I’m) Very sorry. \textbf{In the future}, (I’ll) watch out (about my behavior).                         & \texttt{Pronoun}, \texttt{Ppron}, \texttt{apologetic}, \texttt{Focusfuture} \\ \hline
        		\end{tabularx}
        		\par\end{centering}
        	\caption{\label{tab:weibo-examples}A sample of Weibo and Twitter posts with their English translations (if applicable), with some keywords marked in bold. A selection of related (im-)politeness features are also provided.}
        \end{table*}

        Notice that using in-group markers that are too intimate may suggest impoliteness. Consider the Example 4 in Table \ref{tab:weibo-examples}: the slang word `dawg' steered this Twitter post towards a casual tone of speech, earning this post a borderline standardized politeness score of $0.797$.
        
        One may point out that the feature \texttt{We} has weaker correlations on the Weibo side. This might be explained by the fact that the Chinese language distinguishes between exclusive and inclusive first-person plural pronouns. That is to say, while some pronouns captured by the feature \texttt{We} can be a in-group marker, some other pronouns can be a denial of in-group identity.
        
        \paragraph{Temporal focus.}
        According to Hofstede's Model of Cultural Dimensions, Chinese people adopt a strong long term orientation compared to other cultures~\cite{hofstede2009geert}. This may explain Chinese people's tendency to endure immediate suffering in exchange of a long-term interest\cite{shi2011cultural}. Such difference is reflected in the use of temporal-focusing words: \texttt{Focusfuture} correlates with politeness on Weibo.  In addition, \texttt{Focusfuture} may be used as a replacement for \texttt{Promise}. For example, the Example 5 in Table \ref{tab:weibo-examples} implies a promise without using phrases such as `I promise' and `I will never'. This may explain the insignificance of correlations of \texttt{Promise} in Twitter in Figure \ref{fig:pcc-add}.

\section{RQ3: Predicting Politeness in Chinese and English Microblogs}
    
    Next, we examine how accurately politeness can be predicted in the microblog corpora using machine learning. Pretrained politeness prediction models for both Mandarin and English languages are available at \url{https://github.com/tslmy/politeness-estimator}. 
    
    Aligning with prior work~\cite{danescu-niculescu-mizil2013acomputational}, we use the posts in extreme quartiles of politeness scores. 
    We train linear SVC models on $2,650$ posts ($1,325$ least and $1,325$ most polite posts) from each corpus. The hyperparameters are chosen from a $5$-fold grid search with domains $C\in \{0.01, 0.05, 0.1, 0.25, 0.5\}$ and $\gamma \in \{0.001, 0.0025, 0.005, 0.01, 0.025, 0.05, 0.1\}$. To evaluate the performance of trained models, we use weighted-average F-1 score.
    Since the Stanford API is trained on English corpora, we also train a SVC model with the features extracted from the Stanford API, now part of ConvoKit~\cite{chang2020convokit}, as a baseline for Twitter and compare with other lexical features.

    \begin{table*}[ht]
        \small
        \begin{centering}
        
        \begin{tabular}{|l|l|r|r|r|r|r|}
        \hline
        \textbf{Corpus}       &   \textbf{Feature Set}     &          \textbf{F1} &    \textbf{Precision} &    \textbf{Recall} &    \textbf{ROCAUC} & \textbf{Accuracy} \tabularnewline\hline\hline
        \multirow{6}{*}{Twitter} & LIWC &    $0.755$ &     $0.761$ &   $0.754$ &   $0.757$ &    $0.754$  \tabularnewline\cline{2-7}
              & EmoLex &      0.6 &     $0.659$ &   $0.583$ &   $0.598$ &    $0.583$  \tabularnewline\cline{2-7}
              & PoliteLex &    $0.721$ &     $0.749$ &   $0.716$ &   $0.731$ &    $0.716$  \tabularnewline\cline{2-7}
              & Stanford API &    $0.661$ &     $0.666$ &    $0.66$ &   $0.662$ &     $0.66$  \tabularnewline\cline{2-7}
              & LIWC + EmoLex &     $0.76$ &     $0.767$ &   $0.759$ &   $0.763$ &    $0.759$  \tabularnewline\cline{2-7}
              & LIWC + PoliteLex &   $\textbf{0.772}$ &    $\textbf{0.779}$ &  $\textbf{0.771}$ &  $\textbf{0.775}$ &   $\textbf{0.771}$  \tabularnewline\cline{2-7}
              & EmoLex + PoliteLex &    $0.722$ &     $0.751$ &   $0.717$ &   $0.733$ &    $0.717$  \tabularnewline\cline{2-7}
              & LIWC + EmoLex + PoliteLex &    $0.768$ &     $0.776$ &   $0.767$ &   $0.772$ &    $0.767$  \tabularnewline\hline\hline
        \multirow{6}{*}{Weibo} & LIWC &    $0.729$ &     $0.746$ &   $0.726$ &   $0.736$ &    $0.726$  \tabularnewline\cline{2-7}
              & EmoLex &    $0.597$ &     $0.777$ &   $0.532$ &   $0.563$ &    $0.532$  \tabularnewline\cline{2-7}
              & PoliteLex &    $0.779$ &     $0.804$ &   $0.776$ &   $0.792$ &    $0.776$  \tabularnewline\cline{2-7}
              & LIWC + EmoLex &    $0.743$ &      $0.76$ &   $0.741$ &   $0.751$ &    $0.741$  \tabularnewline\cline{2-7}
              & LIWC + PoliteLex &    $0.803$ &    $\textbf{0.824}$ &   $0.801$ &  $\textbf{0.815}$ &    $0.801$  \tabularnewline\cline{2-7}
              & EmoLex + PoliteLex &    $0.778$ &     $0.805$ &   $0.774$ &   $0.792$ &    $0.774$  \tabularnewline\cline{2-7}
              & LIWC + EmoLex + PoliteLex &   $\textbf{0.804}$ &     $0.822$ &  $\textbf{0.802}$ &   $0.815$ &   $\textbf{0.802}$  \tabularnewline\hline
        \end{tabular}

        \par\end{centering}
        \caption{\label{tab:report}Performance of various lexicon-based feature sets in predicting politeness on Twitter and Weibo corpora. Highest value in each corpus and each metric is marked in bold.}
    \end{table*}

    The performances of lexical features are listed in Table \ref{tab:report}. 
    Lexical features outperform bi-grams and those derived from parse trees (Stanford API) from prior work~\cite{danescu-niculescu-mizil2013acomputational}. Predicting politeness using LIWC and PoliteLex together on Twitter and with the addition of EmoLex on Weibo achieve the highest performance in their corresponding corpora. In general, lexical approach for politeness prediction is seen to generalize across domains.

\section{Conclusion}

    In this paper, we showed that PoliteLex, combined with LIWC and EmoLex, can outperform the Stanford API~\cite{danescu-niculescu-mizil2013acomputational} in predicting politeness on social media corpus and the Stanford politeness corpus~\cite{danescu-niculescu-mizil2013acomputational}. We also studied the similarities and differences in politeness expressions across English Twitter and Mandarin Weibo and found several categories of PoliteLex to be significantly associated with politeness on both platforms. Several correlations mapped onto known psycho-social differences across US and Chinese cultures. For example, use of all pronouns (except honorifics) highly correlates with impoliteness in Weibo, echoing the fact that Chinese is a pro-drop language, one that frequently omits pronouns~\cite{hsiao2014agent}. On Weibo, future-focusing conversations, identifying with a group affiliation, and gratitude are considered to be polite compared to English Twitter. Death-related taboo topics and informal language are associated with higher impoliteness on Mandarin Weibo compared to English Twitter.
    
    
    \paragraph{Limitations and Future Work.} This cross-lingual work only offers preliminary insights based on a relatively small annotated corpora of Weibo and Twitter posts, though comparable to other annotated corpus ~\cite{preoctiuc2016modelling}. Even though the annotators were native language speakers, only two annotators rated each post from the two platforms. Further work is required to refine and validate PoliteLex to produce stable results on larger and more semantically diverse corpora. PoliteLex focuses on semantically unambiguous expressions of (im-)politeness, yet future work could also address semantically ambiguous expressions such as mock politeness (e.g., sarcasm~\cite{taylor2015beyond}). Similarly, in several cultures, the syntactic structure of sentences~\cite{portner2019speaker} may indicate (im-)politeness and could be addressed by future work. Even though we attempted to create an equivalent lexicon based on language translation, since conceptually there is a difference between language and culture, the frequencies may still differ because of culture ~\cite{ji2004culture}. Lastly, although many features considered are seen to map onto prior psychological or pragmatic politeness research, it would be interesting to study causal mechanisms underlying cultural differences and also understand the role of context using word embeddings~\cite{chi2018bcws}.
    
    This study shows the promise of using computational tools for both quantitative and cross-cultural pragmatics. Future work could explore politeness strategies relevant to the current style of communications in the internet age, to enhance PoliteLex while preserving its cross-lingual nature. Future work could also include the annotation of existing cross-lingual datasets~\cite{conneau2018xnli} with politeness information to study translation-relationship between the English and Chinese politeness strategies to inform politeness-aware translation techniques.
\bibliographystyle{ACM-Reference-Format}
\bibliography{paper}

\end{CJK*}
\end{document}